\documentclass[prl,twocolumn,showpacs]{revtex4}
\usepackage{graphicx}
\usepackage{dcolumn}
\usepackage{amsmath}
\usepackage{mathptmx}
\begin{document}

\title{Radiative strength function in $^{96}$Mo reanalyzed}
\author{M.~Guttormsen,$^{1}$\footnote{Electronic address: magne.guttormsen@fys.uio.no} R.~Chankova,$^{2}$ 
U.~Agvaanluvsan,$^{2}$ E.~Algin,$^{5}$ L.A.~Bernstein,$^{2}$ 
F.~Ingebretsen,$^{1}$ T.~L{\"o}nnroth,$^{6}$ S.~Messelt,$^{1}$ G.E.~Mitchell,$^{3,4}$ 
J.~Rekstad,$^{1}$ A.~Schiller,$^{7}$ S.~Siem,$^{1}$ A.C.~Larsen,$^{1}$ A.~Voinov,$^{7,8}$ and S.~{\O}deg{\aa}rd$^{1}$}
\address{$^1$ Department of Physics, University of Oslo, N-0316 Oslo, Norway}
\address{$^2$ Lawrence Livermore National Laboratory, L-414, 7000 East Avenue, Livermore, CA 94551, USA}
\address{$^3$ North Carolina State University, Raleigh, NC 27695, USA}
\address{$^4$ Triangle Universities Nuclear Laboratory, Durham, NC 27708, USA}
\address{$^5$ Department of Physics, Osmangazi University, Meselik, Eskisehir, 26480 Turkey}
\address{$^6$ Department of Physics, {\AA}bo Akademi, FIN-20500 Turku, Finland}
\address{$^7$ Department of Physics and Astronomy, Ohio University, Athens, Ohio 45701, USA}
\address{$^8$ Frank Laboratory of Neutron Physics, Joint Institute of Nuclear Research, 141980 Dubna, Moscow region, Russia}

\begin{abstract}
The radiative strength functions of $^{96}$Mo have been reanalyzed.  
The enhanced $\gamma$ strength for $E_{\gamma}<3-4$ MeV is confirmed. 
\end{abstract}


\maketitle

\section{Introduction}

The Oslo group has reported unexpected enhancements 
in the radiative strength functions (RSF) of low energy $\gamma$-rays for the $^{44,45}$Sc, $^{50,51}$V, $^{56,57}$Fe and $^{93-98}$Mo 
nuclei~\cite{larsen1,larsen2,voinov,Gutt1}. These results, which have been obtained
with the Oslo method~\cite{schi0}, have gained broad interest in the nuclear physics community.
The $^{96}$Mo nucleus has become a benchmark for other experimental 
groups trying to verify or falsify the Oslo results. 
Unfortunately, $^{96}$Mo was not analyzed in an optimal way, 
and we have therefore decided to reanalyze these data sets.  

\section{Reanalyzed data and results}

The two data sets were recorded from the $^{96}$Mo($^3$He,$^3$He$^{\prime}\gamma$)$^{96}$Mo
and $^{97}$Mo($^3$He,$\alpha \gamma$)$^{96}$Mo reactions. In this work we use the same key parameters for the normalizations as previously~\cite{Gutt1,Rositsa}, namely a level density of $\rho(B_n)= 71800$~MeV$^{-1}$ and an average total radiative width of $\langle \Gamma (B_n) \rangle= 150$ meV (from $s$-wave resonances) at the neutron binding energy $B_n$. Experimental details and further references are found in Refs.~\cite{Gutt1,schi0,Rositsa}.

The new analyzes concern two major points.
In the old extraction of RSF, we included the $\gamma$-ray energies 
close or below the strong 778 keV $2^+ \rightarrow 0^+$ ground band transition. 
This transitional region in the experimental $(E_{\gamma},E)$ matrix is not properly subtracted in the first-generation procedure~\cite{Gutt2} and is now excluded.

The second point concerns the estimate of the $\gamma$-ray multiplicity as function of 
excitation energy, which is an important quantity both for normalizing the $\gamma$ spectra
with respect to each other and weighting the higher-order generation spectra in the
subtraction procedure~\cite{Gutt2}. In the previous analyses, we estimated 
the statistical multiplicity at excitation energy $E$ by introducing 
a lower $\gamma$-ray threshold
$E_0$ and an effective excitation energy $E-E_{\rm entry}$ giving
\begin{equation}
\langle M^{\rm stat}_{\gamma}\rangle =(E-E_{\rm entry})/\langle E_{\gamma} \rangle _{> E_0},
\end{equation}
where $\langle E_{\gamma} \rangle _{> E_0}$ is the average energy 
of the $\gamma$ spectrum for $E_{\gamma} > E_0$.
The $E_{\rm entry}$ parameter mimics the excitation energy at which the statistical $\gamma$ transitions enter the ground band. This treatment is applicable to rare earth nuclei, where the detector system efficiency for
the lowest ground state band transitions, typically the $4^+ \rightarrow 2^+$
and the $2^+ \rightarrow 0^+$ transitions, is low. However, for $^{96}$Mo
the energy of the lowest ground band transitions are detected easily. Therefore, in the present analysis we use the straightforward expression for the total $\gamma$-ray multiplicity 
\begin{equation}
\langle M^{\rm tot}_{\gamma}\rangle =E/\langle E_{\gamma} \rangle,
\end{equation}
where we simply divide the excitation energy by the average energy 
of the $\gamma$ spectrum.

\begin{figure}[htb]
\includegraphics[totalheight=9cm]{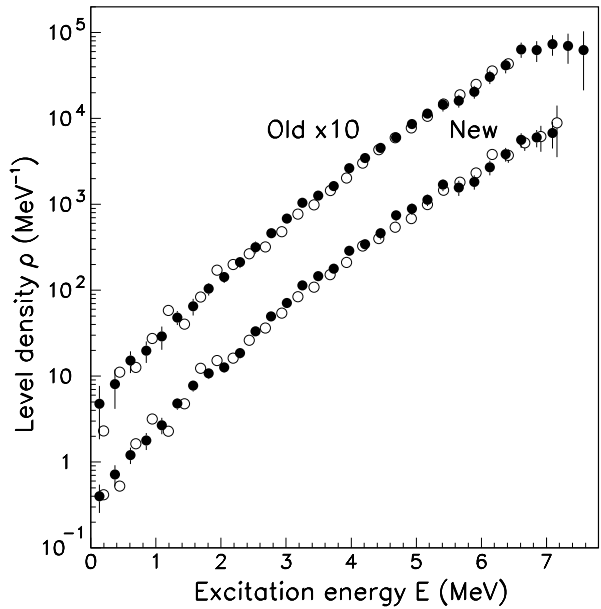}
\caption{Experimental level densities from the  ($^3$He,$\alpha$) (filled circles) and the ($^3$He,$^3$He') (open circles) reaction. The data from the new analysis is compared with previously published data~\cite{Rositsa}.}
\label{fig:fig1}
\end{figure}

\begin{figure}[b]
\includegraphics[totalheight=9cm]{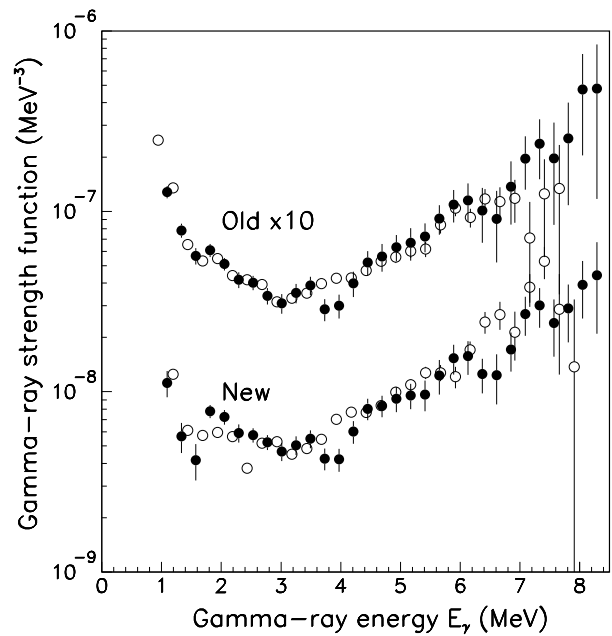}
\caption{Experimental radiative strength functions from the ($^3$He,$\alpha$) (filled circles) 
and the ($^3$He,$^3$He') (open circles) reaction. The data from the new analysis is compared with previously published data~\cite{Gutt2}.}
\label{fig:fig2}
\end{figure}

Both the level density and the radiative strength function will be slightly modified by the new multiplicity expression. Figures~\ref{fig:fig1} and ~\ref{fig:fig2} compare the new level densities and radiative strength functions with previous data \cite{Gutt1,Rositsa}. We see a very good resemblance between the pick-up reaction and the inelastic scattering reaction. The error bars include statistical errors only. The new level densities are very similar to the previous ones. The same is true for the RSFs, except that the upbend is less pronounced due to the exclusion of the 778 keV transition. The values of the data points can be found at http://ocl.uio.no/compilation/.

There have been some misunderstandings concerning the description of the RSF upbend for low $\gamma$ energies given by~\cite{Gutt2}
\begin{equation}
f_{\rm upbend}=\frac{1}{3\pi^2\hbar^2c^2}{\cal{A}}E_{\gamma}^{-b},
\end{equation}
where $\cal{A}$ and $b$ are fit parameters, and $E_{\gamma}$ is given in MeV.
The formula was chosen in order to fit the low-energy data by only two parameters. We would like to stress that this description
should not be used for $\gamma$ energies lower than the experimental data points. In the extreme case when $E_{\gamma} \rightarrow 0$, it is clear that the description is totally unrealistic as it gives wrong $\gamma$ multiplicity.

\section{Summary and conclusions}
The radiative strength function of $^{96}$Mo has been reanalyzed giving a slightly less
pronounced enhancement at lower $\gamma$-ray energies. 
The data points at and below the 778 keV $2^+ \rightarrow 0^+$ transition have been omitted. 
Since extraction of level density is coupled to the radiative strength function, new level densities have also been presented.

\acknowledgments
Financial support from the Norwegian Research Council (NFR) is gratefully acknowledged. Part of this work was 
performed under the auspices of the U.S. Department of Energy by the University
of California, Lawrence Livermore National Laboratory under Contract 
W-7405-ENG-48.  A.V. E.A, U.A, and G.E.M acknowledge support from the National Nuclear Security
Administration under the Stewardship Science Academic Alliances program
through DOE Research Grants No. DE-FG03-03-NA00074, No. DE-FG03-03-NA00076 and U.S.
Department of Energy Grant No. DE-FG02-97-ER41042.

\vspace{\baselineskip}

\end{document}